\documentstyle[emulateapj,psfig]{article}
\submitted{Received 2002 August 21; Accepted: 2002 October 8}

\def\gs{\mathrel{\raise0.35ex\hbox{$\scriptstyle >$}\kern-0.6em
\lower0.40ex\hbox{{$\scriptstyle \sim$}}}}
\def\ls{\mathrel{\raise0.35ex\hbox{$\scriptstyle <$}\kern-0.6em
\lower0.40ex\hbox{{$\scriptstyle \sim$}}}}

\newenvironment{inlinefigure}{%
\def\@captype{figure}%
\noindent\begin{minipage}{0.999\linewidth}\begin{center}\small}
{\end{center}\end{minipage}\smallskip}
\makeatother

\lefthead{Smail et al.}
\righthead{A $z=2.4$ SCUBA Galaxy near 53W002}

\begin{document}

\title{A SCUBA Galaxy in the Protocluster around 53W002 at $z=2.4$}

\author{
Ian Smail,\altaffilmark{1} R.\,J.\ Ivison,\altaffilmark{2} D.\,G.\
Gilbank,\altaffilmark{1} J.\ S.\ Dunlop,\altaffilmark{3} W.\,C.\
Keel,\altaffilmark{4}\\ K.\ Motohara\altaffilmark{5} \& J.\,A.\ Stevens\altaffilmark{2}}

\altaffiltext{1}{Institute for Computational Cosmology, University of Durham, South Road,
        Durham DH1 3LE UK}
\altaffiltext{2}{Astronomy Technology Centre, Royal Observatory, 
        Blackford Hill, Edinburgh EH9 3HJ UK}
\altaffiltext{3}{Institute for Astronomy, University of Edinburgh, 
        Blackford Hill, Edinburgh EH9 3HJ UK}
\altaffiltext{4}{Department of Physics \& Astronomy, University of Alabama,
        Tuscaloosa, AL 35487 USA}
\altaffiltext{5}{Institute of Astronomy, University of Tokyo, Mitaka, Tokyo 181-0015 Japan}

\setcounter{footnote}{4}

\begin{abstract}
We analyse an 850-$\mu$m SCUBA map of the environment of the $z=2.39$
radio galaxy 53W002, which has been shown to reside in an over-density
of Ly$\alpha$-detected galaxies.  We identify four luminous
submillimeter (submm) sources within a 2.3$'$ (1.2\,Mpc at $z=2.39$)
diameter area around the radio galaxy (which itself is a weak submm
source).  We employ a 1.4-GHz map to accurately locate the counterpart
of one of these sources, SMM\,J17142+5016, and identify this source
with a narrow-line AGN with an extended Ly$\alpha$ halo at $z=2.390$
which is member of the structure around 53W002.  Hence SMM\,J17142+5016
is the first spectroscopically-confirmed, submm-selected companion to a
high-redshift radio galaxy.  We discuss the OHS $JHK$ spectrum of this
galaxy and in addition present five new constraints on its spectral
energy distribution long-ward of 1\,$\mu$m, using these to estimate its
bolometric luminosity as $\sim 8\times 10^{12}$\,L$_\odot$, or a star
formation rate of $\sim 10^3$\,M$_\odot$\,yr$^{-1}$ if young stars
provide the bulk of the luminosity.  This result provides direct
support for the statistical detection of over-densities of SCUBA
galaxies around high-redshift radio galaxies and confirms theoretical
predictions that SCUBA galaxies, as the progenitors of massive
ellipticals, should be strongly clustered in the highest density
regions of the distant Universe.
\end{abstract}


\section{Introduction}

%
%
\begin{figure*}[tbh]
\centerline{\psfig{file=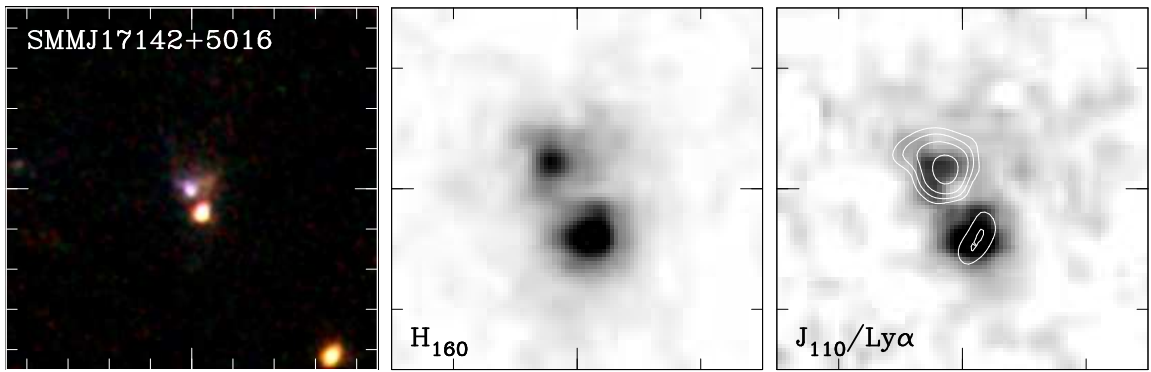,width=6.0in,angle=0}}
\caption{\small Three images demonstrating the varied morphology of
SMM\,J17412+5016\#18 in the rest-frame optical/UV.  These show (from
left to right): a true-color representation constructed from the {\it
HST}/WFPC2 $B_{450}V_{606}I_{814}$ imaging; an expanded view of the
$H_{160}$ image to better illustrate the morphology of the galaxy in
the rest-frame optical; and a view which contrasts the morphology of the
$J_{110}$-band, rest-frame near-UV continuum emission (shown as a
gray-scale) with the morphology in the F410M filter which is dominated
by the Ly$\alpha$ emission and is shown as a logarithmic contour
plot. ~From these images we see that the optical counterpart to
SMM\,J17412+5016 comprises a compact red component to the South and a
more diffuse blue structure to the North-East with an extension to the
West.  At $z=2.39$, the $K$-band magnitude of SMM\,J17142+5016\#18
corresponds to an apparent magnitude of $M_V\sim -23$, however, its
present-day luminosity depends critically on the competing effects of
current dust extinction and subsequent star formation and hence the
$z=0$ absolute luminosity of this galaxy is extremely uncertain.  The
left-hand panel is $9''\times 9''$ (70\,kpc at $z=2.39$), while the
central and right-hand panels are $3''\times 3''$ each, all have North
top and East left. }
\end{figure*}

The extreme luminosities exhibited by 
the dusty galaxy population which contributes
the bulk of the extragalactic background in the far-infrared and submm
has emphasized the importance of dust-obscured activity in the early
evolution of massive galaxies and black holes (Blain et al.\ 1999a;
Cowie et al.\ 2002).  The majority of the dusty, active systems
detected by SCUBA appear to lie at $z>1$, their typical bolometric
luminosities are $10^{11}$--10$^{13}$\,L$_\odot$ and their space
densities are $\sim 10^{-4}$\,Mpc$^{-3}$, 2--3 orders of magnitude
higher than similar luminosity galaxies at $z\sim 0$, indicating
strong evolution of this population (Smail, Ivison \& Blain 1997;
Lilly et al.\ 1999; Blain et al.\ 1999b; Cowie et al.\ 2002;
Chapman et al.\ 2002b).
However, little is known about the types of environment which these
galaxies inhabit.  Similar luminosity, dusty systems in the local
Universe are typically found in low-density regions and they avoid the
dense cores of rich clusters of galaxies (Tacconi et al.\ 2002).  The
situation is expected to be very different at high redshifts where,
if they truly represent the progenitors of massive ellipticals, the
SCUBA galaxies should be clustered around the highest density regions
which will subsequently evolve into the cores of rich clusters
at the present-day (Ivison et al.\ 2000b, I00b).

Due to the small samples available in the submm waveband observational
evidence of this clustering is tentative at the present-time (Scott et
al.\ 2002; Webb et al.\ 2002).  Nevertheless, there are suggestions
that SCUBA galaxies are strongly clustered, in particular associations
of SCUBA galaxies with other classes of clustered high-redshift
sources, such as Lyman-break galaxies or X-ray sources, have been
serendipitously found (Chapman et al.\ 2001; Ledlow et al.\ 2002;
Almaini et al.\ 2002).  A more direct approach to tackle this question
has been undertaken by I00b in a targeted survey of regions around
high-redshift, powerful AGN.  These are expected to preferentially
inhabit high-density regions in the early Universe (e.g.\ West 1994), a
prediction which has some observational support from the discovery of
excesses of both Extremely Red Objects and Ly$\alpha$ emitters around
some high-redshift radio galaxies (e.g.\ Arag\'on-Salamanca et al.\
1994; Lacy \& Rawlings 1996; Pascarelle et al.\ 1996a; Yamada et al.\
1997; Keel et al.\ 1999; Venemans et al.\ 2002).  The first results
from the I00b survey show a significant over-density of SCUBA galaxies
around the signpost high-redshift AGN, but confirming this association
with redshifts for the SCUBA sources is very demanding.

One of the fields covered in the I00b survey is that surrounding the
$z=2.39$ steep-spectrum, narrow-line radio galaxy 53W002 (Windhorst et
al.\ 1991, 1998).  This region is especially interesting as it has been
shown to contain an over-density of compact, Ly$\alpha$ emission-line
galaxies at $z\sim 2.4$ (Pascarelle et al.\ 1996a, P96a, 1998, P98; Keel
et al.\ 1999, K99).  Nine of these emission-line galaxies have been
spectroscopically confirmed as companions to the radio galaxy, with a
velocity dispersion of $\sim 400$\,km\,s$^{-1}$ and a spatial extent of
$\sim 4$\,Mpc (K99).  In this paper we report on the SCUBA observations
of the 53W002 field which uncover four luminous, submm galaxies. By
matching the submm source position using an astrometrically-precise
1.4-GHz map we show that one of these sources is coincident
with a Ly$\alpha$-selected galaxy at $z=2.39$, confirming the presence
of ultraluminous, dusty galaxies in the over-dense structure around
53W002 at a look-back time of 11\,Gyrs (we adopt a cosmology with
$q_o=0.5$, and H$_{\rm o}=50$\,km\,s$^{-1}$\,Mpc$^{-1}$).

We discuss new and archival observations of the 53W002 field
in the next section, present our analysis and results in \S3,
discuss these in \S4 and give our conclusions in \S5.

\section{Observations and Reduction}

\subsection{Submm Mapping}

We observed a $\sim 2.3'$-diameter field centered on 53W002 at 450-
and 850-$\mu$m during 2001 March 3--6 using the SCUBA bolometer array
(Holland et al.\ 1999) on the James Clerk Maxwell Telescope
(JCMT)\footnotemark. The conditions were good and stable for the four
nights when data were taken, with an 850-$\mu$m opacity measured every
hour of 0.12--0.23 for the first three nights, rising to 0.35 on the
final night. Flux calibration used beam maps of Mars and the nightly
calibration factors were consistent at the 5\% level (r.m.s.)  at
850\,$\mu$m.  

To map the field, the secondary mirror followed a jiggle
pattern designed to fully sample the image plane, chopping East-West
by 30$''$ at 7\,Hz whilst the telescope nodded between the same
positions every 16\,s in an on--off--off--on pattern.  The total
exposure time was 33.3\,ks.  The 450- and 850-$\mu$m maps were created
using {\sc surf} (Jenness \& Lightfoot 1997). In order to extract
reliable source positions and flux densities, the maps were
deconvolved using the symmetric $-1, +2, -1$ zero-flux beam (which
arises from chopping and nodding) and a modified version of the {\sc
clean} algorithm (H\"ogbom 1974) as described by I00b.  This process
produces a restored map at the native resolution of the JCMT, 14$''$
FWHM, but without the negative side-lobes from the chopping of bright
sources.

\footnotetext{The JCMT is operated by the Joint Astronomy Centre on
behalf of the United Kingdom Particle Physics and Astronomy Research
Council (PPARC), the Netherlands Organisation for Scientific Research,
and the National Research Council of Canada.}

%
%
\begin{figure*}[tbh]
\centerline{\psfig{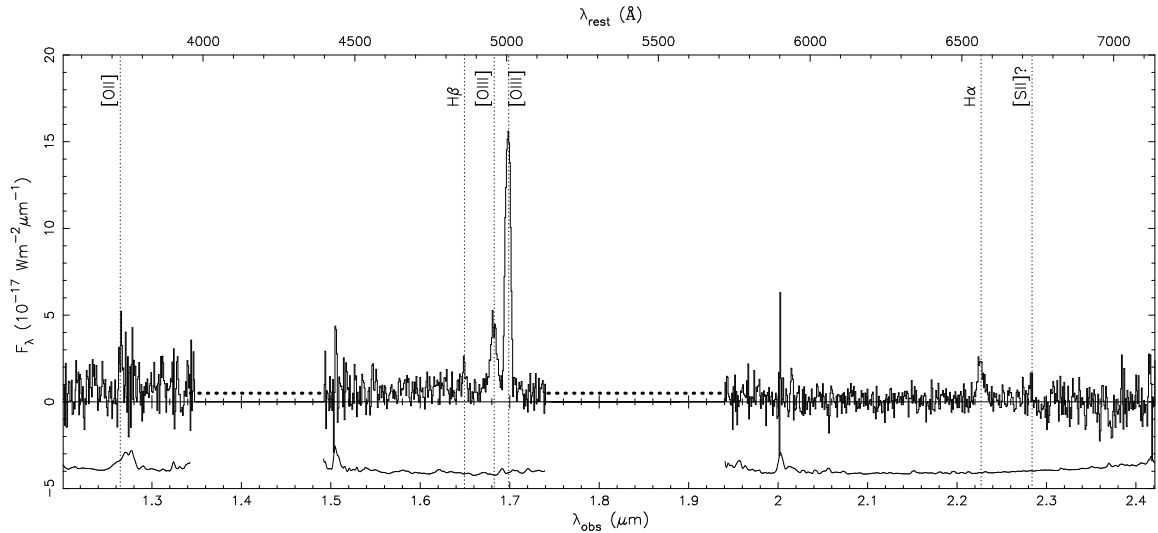}}
\caption{\small 
The combined $JHK$ OHS spectrum of SMM\,J17142+5016/\#18, we identify a
number of prominent restframe-optical emission lines, note 
the relative strength of the [O{\sc iii}]\,$\lambda$5007 line
indicating the galaxy is likely to host a Seyfert-2 nucleus. The top axis
gives the restframe wavelength of these features, while the bottom axis
is in the observed frame. The lower line shows the sky spectrum and
the heavy dotted lines indicate regions of the spectrum which are
dominated by strong atmospheric absorption. }
\end{figure*}

\subsection{Radio Mapping}

The radio map of the 53W002 field from the National Radio Astronomy
Observatory's (NRAO) Very Large Array (VLA),\footnote{NRAO is operated
by Associated Universities Inc., under a cooperative agreement with the
National Science Foundation.} is dominated by the 50-mJy central radio
source (Windhorst et al.\ 1991, 1998) and seven bright, nearby
sources. The presence of these bright sources makes self-calibration of
the radio data trivial, but ultimately limits the dynamic range of the
map in the vicinity of the radio source.  

For these observations we employed B configuration, obtaining a total
of 10.8\,ks of useful integration on the 53W002 field during 2001 May 14
and May 17.  We scanned the amplitude/phase calibrator, 1725+455,
every hour and the flux scale was tied to 0137+331.  To ameliorate
the effects of band-width smearing the data were taken and reduced in
pseudo-continuum, spectral-line mode, using 28 3.25-MHz
dual-polarization channels centered at 1.4\,GHz, with data recorded
every 5\,s.  After editing and standard calibration of the data (and
their associated weights), using {\sc aips}, the {\sc imagr} task was
used to map a $34' \times 34'$ field, with simultaneous imaging of 16
bright nearby sources identified from the NRAO VLA Sky Survey (NVSS,
Condon et al.\ 1998).  Using these maps, made with {\sc robust = 0}
weighting of the visibilities, {\sc clean} boxes were positioned
around obvious sources, and {\sc imagr} was re-run with 5,000
iterations of the {\sc clean} algorithm (H\"ogbom 1974). The
resulting {\sc clean} components were then used as a model for self
calibration (in phase only).  The mapping was repeated, after checks
on the {\sc clean} boxes, followed by a further self-calibration step
to produce the final map. After correction for the primary beam
response of the VLA antennas using {\sc pbcor}, the resulting map has
a noise level of 30\,$\mu$Jy\,beam$^{-1}$, with a $5.3'' \times 3.7''$
beam.

\subsection{Optical and Near-infrared Imaging}

Our analysis also exploits archival {\it Hubble Space Telescope} ({\it
HST}) imaging of the 53W002 field in both the optical (WFPC2) and
near-infrared (NICMOS).  We retrieved re-calibrated and reduced  
WFPC2 imaging of the field from the {\it HST} archive at
ST-ECF\footnote{Based on observations made with the 
{\it Hubble Space Telescope} obtained from the ESO/ST-ECF Science Archive
Facility.}, comprising three broad-band exposures: 57.6\,ks in F450W
($B_{450}$), 27.2\,ks in F606W ($V_{606}$) and 20.4\,ks in F814W
($I_{814}$), as well as a 40.5-ks exposure in the F410M medium band.
The latter filter serves to isolate the Ly$\alpha$ emission line at
$z\sim 2.4$.  More information on these data is given in P96a and P98.

The acquisition and reduction of the NICMOS imaging of this field, in
the F110W ($J_{110}$) and F160W ($H_{160}$) filters, are described in
detail by Keel et al.\ (2002, K02).  A total of 2.4-ks integration was
obtained in each filter on two fields around 53W002 using the NIC3
camera during the special NIC3 campaign in January 1999.  These
observations have a resolution of 0.2$''$ FWHM and  3-$\sigma$ point-source
sensitivities of $J_{110}\sim 24.3$ and $H_{160}\sim 23.9$.  

While the NICMOS images have exquisite resolution, they cover only a
small proportion of the SCUBA map and so we have also obtained
wider-field $K_s$-band imaging covering the full map.  These data were
taken with the INGRID near-infrared imager (Packham et al.\ 2002) on
the 4.2-m William Herschel Telescope (WHT)\footnotemark.  Our
observations were obtained on the night of 2001 May 10 and consist of a
total of 4.3\,ks integration in the $K_s$ filter in photometric
conditions and $0.6''$ seeing.  The reduction and calibration of these
data uses the pipeline described in Gilbank et al.\ (2002).

\footnotetext{Based on observations made with the William Herschel
Telescope operated on the island of La Palma by the Isaac Newton Group
in the Spanish Observatorio del Roque de los Muchachos of the
Instituto de Astrofisica de Canarias.}
\bigskip

\subsection{Near-infrared Spectroscopy}

Near-infrared spectroscopy of several galaxies in the field of 53W002,
in particular object \#18, was obtained by Motohara et al.\
(2001a,b). These observations used the newly commissioned OH-airglow
Suppression Spectrograph (OHS; Iwamuro et al.\ 2001) and Cooled
Infrared Spectrograph and Camera for OHS, CISCO (Motohara et al.\ 2002)
on the 8.2-m Subaru Telescope.\footnote{Based on data collected at
Subaru Telescope, which is operated by the National Astronomical
Observatory of Japan.}  More details of the observations are given in
Motohara et al.\ (2001b).  The spectra cover the $JHK$ bands in two
configurations, in both instances the slit was placed at a PA of $-$35
degrees and centered on the redder, southern component in the galaxy
(Fig.~1).

The $JH$-band spectrum of \#18 was obtained on the night of 1999 May
21, using OHS in the light path in front of CISCO to suppress the
atmospheric OH emission.  A total of 8\,ks of data was acquired in
$0.6''$ seeing as a series of 1\,ks exposures, with the $5''$ nods
along the slit between exposures.  Using a $0.95''$-wide slit the
resolution was $\sim 200$ in the middle of the $H$-band.  Atmospheric
transmission corrections came from observations of SAO\,30245 and
wavelength calibration was obtained from the OH lines measured in
observations without OHS.  Checks of the instrument stability suggest
that this calibration is good to $<5$\,\AA.

The $K$-band spectroscopy was obtained on 1999 May 3--4 with just
CISCO, at the Cassegrain focus, in $0.4''$ seeing and non-photometric
conditions.  The $0.7''$ slit resulted in a resolution of $\sim430$ at
$2.2\mu$m.  A total of 4.8\,ks of integration in $24\times 200$\,s
exposures was obtained, nodding the telescope by $\sim 5''$ after every
6 exposures to allow for sky subtraction.  Calibration of the
atmospheric transmission came from observations of SAO\,30082 taken
immediately before the science frames on 1999 May 4 and wavelength
calibration used the sky lines.

The spectroscopic data were reduced in a standard manner, including
flat-fielding, sky subtraction, correction of bad pixels and residual
sky subtraction. The spectra of the SAO stars were used to correct for
atmospheric extinction.  As the seeing was smaller than the slit width
for both observations we adopt a quarter of the slit width as our
estimate of the absolute calibration error on our wavelength scale.
This amounts to 22\,\AA\ in the $JH$-band spectrum and 13\,\AA\ in the
$K$-band.  The spectra were flux calibrated using the broad-band
photometry of the source from Yamada et al.\ (2001) and Motohara et
al.\ (2001b).  A $1.2''\times 2.1''$ aperture, aligned with the slit
was used in both cases.  The final spectrum covering the $JHK$ bands is
shown in Figure~2 and an initial analysis of this spectrum is described
in Motohara et al.\ (2001a).

\section{Analysis and Results}

We detect four significant sources in the 850-$\mu$m SCUBA map of the
53W002 field above a 4-$\sigma$ limit of 3.7\,mJy, this represents a
modest excess over the blank-field expectation of 2 sources at this
flux limit.\footnote{We also detect emission from 53W002 at lower
significance, with an 850-$\mu$m flux of $3.1\pm 0.9$\,mJy, consistent
with the 3-$\sigma$ upper limit of $<3.3$\,mJy from Hughes \& Dunlop
(1998).}  Unfortunately, the large JCMT beam means that it is
impossible to reliably locate the source of this submm emission.
However, by exploiting the smaller beam and better absolute astrometric
precision of our VLA map we can provide a more accurate position for
any submm source detected at 1.4\,GHz. These positions can then be
compared to our near-infrared images, with a precision of $\sim 0.5''$,
to search for optical/near-infrared counterparts.

Only one of the submm sources is detectable at 1.4-GHz (in addition to
53W002) which we denote as SMM\,J17142+5016, with a nominal position
from the 850-$\mu$m map of 17\,14\,12.18 $(\pm 0.16^s)$, +50\,16\,02.8
$(\pm 1.8'')$ (J2000).  This galaxy has an 850-$\mu$m flux of $5.6\pm
0.9$\,mJy and a 3-$\sigma$ limit on the 450-$\mu$m flux of
$<$30\,mJy. It lies 1.3$''$ from a $260\pm 30 \mu$Jy radio source (at
17\,14\,12.040 $(\pm 0.028^s)$, +50\,16\,02.92 $(\pm 0.14'')$) in the
1.4-GHz map.  This radio source is also detected at 8.4-GHz with a flux
of $37\pm 7\mu$Jy (Fomalont et al.\ 2002), indicating a spectral slope
of $\alpha = -1.1$. Using the positions of other radio sources in the
INGRID field we place the radio source within $1.0''$ of an optical
galaxy (nominally at 17\,14\,11.99, +50\,16\,02.1), identified as \#18
in P96a. We conclude that \#18 is the likely counterpart to
SMM\,J17142+5016 and in the following discussion we identify the galaxy
as SMM\,J17142+5016/\#18.

\subsection{SMM\,J17142+5016/\#18}

SMM\,J17142+5016/\#18, is one of the brightest of the $\sim 14$
Ly$\alpha$ emission-line sources detected at $z\sim 2.4$ in a
narrow-band imaging search around 53W002 (P96a; Pascarelle et al.\
1996b; P98; K99; K02; Yamada et al.\ 2001; Motohara et al.\ 2001a,b).
We summarize some of the relevant, published information on this
galaxy here: the {\it HST} imaging of SMM\,J17142+5016/\#18 shows that
it is distinguished by its large angular size, $\sim 2''$, in contrast
to the typically unresolved light profiles of the majority of the
Ly$\alpha$ emitters in this region (P96a).  SMM\,J17142+5016/\#18
consists of pair of components separated by 0.6$''$ (4.7\,Kpc at
$z=2.39$), see Fig.~1. The north-eastern component has bluer colors
and a more extended light profile than the southern component (whose
intrinsic FWHM is only 0.25$''$ in the $I_{814}$-band, indicating a
half-light radius of just 1\,kpc at $z=2.39$).  The restframe UV
spectrum published by P96a shows a strong, but relatively narrow
Ly$\alpha$ line, weak C{\sc iv}\,$\lambda$1549 and possibly N{\sc v}\,$\lambda$1239, confirming its
redshift as $z=2.393$.  In conjunction with the preliminary analysis
of the OHS near-infrared spectrum of this galaxy by Motohara et al.\
(2001a), the spectral properties are very similar to those of a
Seyfert-2.  Deeper, ground-based Ly$\alpha$ imaging of this galaxy by
K99 shows that it also possess a very extended Ly$\alpha$ halo, $\gs
50$\,kpc.
~\medskip

%
%
\hspace*{-0.7cm}{\scriptsize
{\centerline{\sc Spectral Properties of SMM\,J17142+5016/\#18}}
\smallskip

\hspace*{-0.7cm}\begin{tabular}{lrrccl}
\hline\hline
\noalign{\smallskip}
Line      &  $\lambda_{\rm obs}$~  & $\lambda_{\rm o}$~~ & $z^a$ &
Flux & Comments \cr
          & (\AA)~~ & (\AA)~~ &  & ($10^{-19}$\,W\,m$^{-2}$) & \cr
\hline
\noalign{\smallskip}
Ly$\alpha$ & 4124.9 & 1215.7 & 2.393 & 4.7 &  K02 \cr
\noalign{\smallskip}
[O{\sc ii}] & 12651.5 & 3727.4 & 2.394 & $1.2\pm 0.2$ & \cr
H$\beta$   & 16487.3 & 4861.3 & 2.392 & $0.54\pm 0.20$ & \cr
[O{\sc iii}]& 16817.1  & 4959.0 & 2.391 & $3.22\pm 0.10$ & \cr
[O{\sc iii}]& 16975.6 & 5006.9 & 2.390 & $10.5\pm 0.3$ & \cr
H$\alpha$  & 22260.1 & 6562.8 & 2.392 & $1.96\pm 0.05$ & Blended with
[N{\sc ii}] \cr 
  & 22248.9 & 6562.8 & 2.390 & $1.4\pm 0.1$ & Deblended \cr 
\hline
\end{tabular}
\medskip

\hspace*{-0.7cm}\begin{tabular}{l}
$^a$ Systematic uncertainties in the redshifts
from the wavelength calibration\\ are
$\pm$0.004 for the $J/H$-band
 spectra and $\pm$0.002 for the $K$-band lines.\cr
\end{tabular}
}

\subsection{Morphology}

Turning now to the new data presented in this paper, we note that the
morphology of SMM\,J17142+5016/\#18, showing two components separated
by $\sim 1''$ and extended emission, is remarkably similar to that of
other optically-bright SCUBA galaxies, SMM\,J02399$-$0136 (a
narrow-line, BAL QSO at $z=2.80$, Ivison et al.\ 1998),
SMM\,J14011+0252 (a star-burst galaxy at $z=2.56$, Ivison et al.\ 2000a,
2001) and Westphal-MMD11 (a submm-bright Lyman
break galaxy at $z=2.98$, Chapman et al.\ 2002a).  All of these very
luminous SCUBA galaxies also exhibit multiple components, apparent
tidal features and complex emission-line structures on scales of $\sim
10$\,kpc.  We compare the spatial distribution of the high-surface
brightness Ly$\alpha$ emission (from the F410M image) and UV continuum
(given by the $J_{110}$-band, although this includes a small
contribution from [O{\sc ii}]\,$\lambda$3727, Fig.~2) in the right-hand
panel of Fig.~1.  This shows that a large fraction of the Ly$\alpha$
emission from the system arises in the northern component, even though
the underlying continuum in this feature is weak.  This component may
therefore represent either lower-mass companion, tidal debris or
scattered light from an obscured AGN.  Unfortunately the precision of
the relative astrometry between our radio and near-infrared images
precludes us more precisely locating the source of the radio (and hence
submm) emission in this system (c.f.\ Ivison et al.\ 2001).

\subsection{Restframe UV/optical spectral properties}

We identify a number of strong emission lines in the OHS spectrum of
SMM\,J17142+5016/\#18 in Fig.~2, including H$\alpha$, H$\beta$, [O{\sc
ii}]\,$\lambda$3727, [O{\sc iii}]\,$\lambda\lambda$4959, 5007, the
signature of [N\,{\sc ii}]\,$\lambda$6583 blended with H$\alpha$ and
perhaps [S\,{\sc ii}]\,$\lambda$6731.  The emission-line properties
are listed in Table~1.  The redshift for SMM\,J7142+5016/\#18 is
$z=2.390\pm 0.002$ from the H$\alpha$ line, compared to $z=2.393$ from
the peak of the Ly$\alpha$ line (P96a).  This small offset may
indicate the presence of absorption in the blue wing of Ly$\alpha$
line, which would have been missed in the low signal to noise spectrum
shown in P96a.  The small differences between redshifts from different
lines in the {\it JH\/}- and $K$-bands are within the calibration
uncertainties of the observations.  

The OHS spectra show that the H$\alpha$ and [O{\sc iii}] emission is
extended by 1--2$''$ along the slit to the North-West from the compact,
southern component.  In addition the [O{\sc iii}]\,$\lambda$5007 line
shows a velocity gradient of 200--300\,km\,s$^{-1}$ across this
extension.  The relatively low resolution of the OHS observations means
that, apart from the spatial velocity gradient, none of the emission
lines are resolved, with limits of $\sigma \ls 1500$\,km\,s$^{-1}$ for
lines in the {\it JH\/}-band and $\sigma \ls 800$\,km\,s$^{-1}$ for
those in the $K$-band.  As an aside, we note that the galaxy lies
outside the sensitive area of the CO line searches around 53W002 and so
no useful dynamical information is currently available (Scoville et
al.\ 1997; Alloin et al.\ 2000).

The H$\alpha$+[N{\sc ii}] flux measured from the OHS spectrum is
$(1.96\pm 0.05)\times 10^{-19}$\,W\,m$^{-2}$. This compares to
$6.3\times 10^{-19}$\,W\,m$^{-2}$ from the narrow-band imaging of K02,
further strengthening the claim of a significant, spatially extended
component to this line emission, but also underlining the difficulty of
extracting reliable line-ratio estimates from the different published
long-slit measurements.  We estimate the contribution from the blended
[N{\sc ii}]\,$\lambda$6583 line to the H$\alpha$ in the long-slit
spectrum by fitting a model with two Gaussian profiles, assuming both
lines have the same FWHM, to the data.  We derive a line ratio of
[N\,{\sc ii}]\,$\lambda$6583/H$\alpha = 0.4\pm 0.1$, similar to that
seen in normal star-forming galaxies.  However, the low signal to noise of
the H$\beta$ flux measurement precludes any detailed analysis of
the reddening in this system.

The broad-band colors of the SMM\,J17142+5016/\#18 listed in K02 appear
unusual, being blue short-ward of 2500\AA\ in the restframe,
$(B-I)=1.3$, and comparatively red long-ward of this, $(I-K)=3.1$ (K02;
Yamada et al.\ 2001).  The cause of this apparent reddening of the SED
is the large contribution to broad-band fluxes from red-shifted
Ly$\alpha$ and optical emission lines (Fig.~2).  These contributions
amount to 25\% of the light in the $B$-band (Ly$\alpha$), 10\% of the
$J$-band flux ([O{\sc ii}\,$\lambda$3727), 45\% in the $H$-band
(H$\beta$, [O{\sc iii}]\,$\lambda$4959,5007) and 30\% in the $K$-band
(H$\alpha$, [N{\sc ii}]).  Correcting for this line emission we
estimate the colors of the underlying continuum as: $(B-I)=1.6$,
$(I-K)=2.7$, $(J-H)=1.3$ and $(H-K)=0.6$.  These colors are similar to
those expected for a young stellar population at $z=2.4$, although the
absence of strong stellar absorption features in the restframe UV
spectra (P96a; Keel priv.\ comm.) suggests that there is a significant AGN
contribution to the UV emission.

The restframe optical emission line ratios measured from the OHS
spectrum of SMM\,J17142+5016/\#18 also suggest the presence of an AGN.
Motohara et al.\ (2001a) discuss the flux ratios for this galaxy on
diagnostic diagrams for [O{\sc iii}]\,$\lambda$5007/H$\beta$, [O{\sc
ii}]\,$\lambda$3727/[O{\sc iii}]\,$\lambda$5007 and [N\,{\sc
ii}]\,$\lambda$6583/H$\alpha$.  They conclude that the system has
emission line ratios similar to those seen in Seyfert-2's, although a
low-metallicity star-burst is also possible.

~\smallskip

%
%
\begin{inlinefigure}
\centerline{\psfig{file=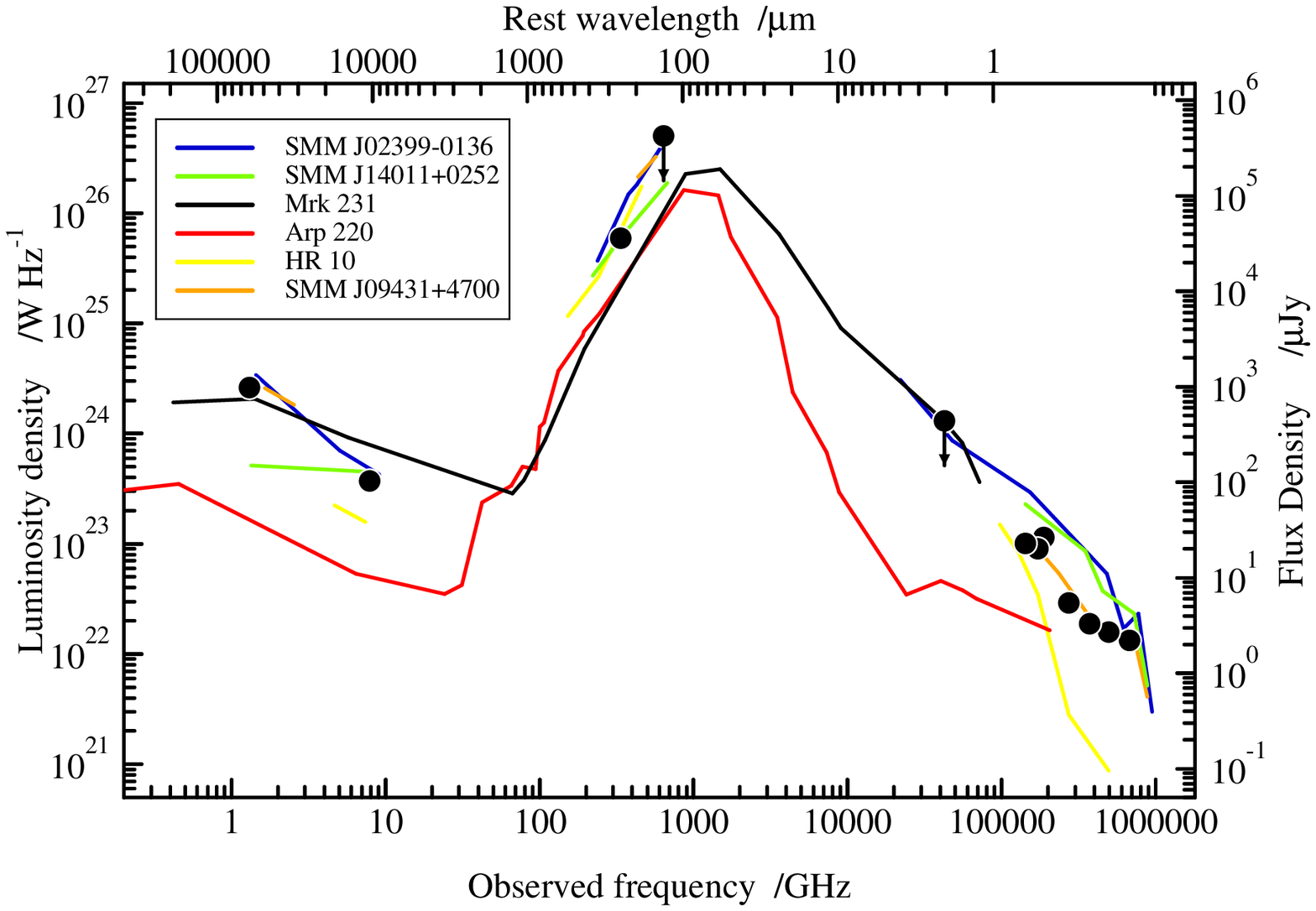,width=3.0in,angle=0}}
\end{inlinefigure}
\vspace*{-3mm}

\noindent{\small\addtolength{\baselineskip}{-2pt} {\sc Fig. 3 ---} A
comparison of the SED of SMM\,J17142+5016/\#18 with a range of
well-studied, luminous dusty galaxies at low and high redshifts.  The
SED appears to very similar to SMM\,J09431+4700, an ultraluminous,
narrow-line Seyfert-1 at $z=3.39$ (Ledlow et al.\ 2002), or a lower
luminosity analog of SMM\,J02399$-$0136 (Ivison et al.\ 1998).  The
restframe optical/UV photometry of SMM\,J17142+5016/\#18 comes from
K02 and we include a 3-$\sigma$ upper limit on the 7\,$\mu$m emission
from the galaxy based on {\it ISO} observations (Keel, priv.\ comm.).

}
\smallskip

\subsection{Far-infrared/radio properties}

In Fig.~3 we compare the spectral energy distribution (SED) of
SMM\,J17142+5016/\#18 to well-studied SCUBA galaxies and other dusty,
luminous galaxies.  The galaxy's SED is typical of ultraluminous
infrared galaxies and shows strong similarities to the $z=3.39$
SCUBA-selected narrow-line Seyfert-1 SMM\,J09431+4700 (Ledlow et al.\
2002).  Taking the 850-$\mu$m flux of SMM\,J17142+5016 and assuming a
dust spectrum with $T_d=38$\,K and $\beta=1.5$, we estimate a dust
mass of $\sim 10^8$\,M$_\odot$, and a bolometric luminosity of $(8\pm
2)\times 10^{12}$\,L$_\odot$, this translates into a star formation
rate for $>5$\,M$_\odot$ stars of $\sim 750$\,M$_\odot$\,yr$^{-1}$
(Condon 1992), or $\sim 4\times 10^3$\,M$_\odot$\,yr$^{-1}$ accounting
for stars with masses of $>0.1$\,M$_\odot$ using a Salpeter IMF.

We can also estimate the star formation rate in SMM\,J17142+5016 from
the H$\alpha$ flux (Table~1; K02), this indicates an H$\alpha$
luminosity of 0.3--$1\times 10^{36}$\,W, suggesting a star formation
rate in the range $10^2$--$10^3$\,M$_{\odot}$\,yr$^{-1}$ (Kennicutt
1983; Barbaro \& Poggianti 1997). The contribution from any
AGN-powered component of the line will bring this estimate down.
Nevertheless, these measurements are far in excess of the estimated
star formation rate based on the UV luminosity of $\sim
10$\,M$_\odot$\,yr$^{-1}$ (not accounting for any AGN contribution,
P96a), indicating a dust-incurred deficit in the UV luminosity which
was already noted as a possible cause of the high Ly$\alpha$
equivalent width (K02; see also Taniguchi \& Shioya 2001).

The radio counterpart to SMM\,J17412+5016 is unresolved at both 1.4-
and 8.4-GHz, which place limits on its FWHM of $\ls 3''$, or $\ls 23$\,kpc at
$z=2.39$.  Using the 1.4-GHz and submm fluxes we can also estimate the
radio-submm spectral index, $\alpha_{850}^{1.4}$ (Carilli \& Yun 1999,
2000), at $\alpha_{850}^{1.4}=0.56 \pm0.06$, which predicts
$z=1.0_{-0.4}^{+0.7}$ based on the Carilli \& Yun (2000) models.  This
is at odds with the spectroscopic redshift of $z=2.390$ (Table~1) and
suggests that there is an additional contribution to the 1.4-GHz
emission from the AGN (as also indicated by the steep radio spectrum
at high frequencies).  It will be important to place a precise limit
on the bolometric luminosity of the AGN using the {\it Chandra} X-ray
observations of this field and so constrain the relative contribution
from the AGN and star formation to the dust emission from this
galaxy (Bautz et al.\ 2000).

\section{Discussion}

The properties of the components of SMM\,J17142+5016/ \#18 hint at a
complex mix of AGN emission, star formation, dust obscuration and
scattered radiation, which is difficult to unravel.  Taking the
southern component first: it is clear that this hosts an AGN, although
there is also extended UV light detected from the host galaxy.
Assuming the line ratios in the OHS spectrum (Fig.~2) are dominated by
this component, then this source should be classified as a Seyfert-2.
In contrast the north-eastern component shows weak continuum and very
strong Ly$\alpha$ emission (Fig.~1) suggesting it may be a lower-mass
star-forming companion or tidal debris which is either locally ionized
or illuminated by scattered light from the AGN.  Support for the latter
explanation comes from the apparent detection of C{\sc iv} emission
across this structure (K99; K02). A definitive
conclusion about the nature of the north-eastern component awaits the
publication of higher-quality, spatially resolved optical spectra of
this structure.

Perhaps the most interesting feature of SMM\,J17142+5016/\#18 is the
extended Ly$\alpha$ halo around this galaxy (K99; K02), which may also
have been seen in H$\alpha$.  This low-surface brightness feature is
on a much larger scale than the high-surface brightness Ly$\alpha$
emission in Fig.~1, extending out east from the galaxy at least $6''$
(50\,kpc). This extended Ly$\alpha$ emission in the same direction as
the extended [O{\sc iii}]\,$\lambda$5007 line discussed earlier, with
the velocity gradient then reflecting motion in the halo gas induced
by a tidal interaction or a wind.  The combination of strong submm
emission and an extended Ly$\alpha$ cloud is somewhat surprising,
given the propensity for dust to absorb Ly$\alpha$ photons, although
it has been seen before (Chapman et al.\ 2001, 2002b). Taniguchi \& Shioya
(2001) have suggested that this phenomenon arises from an extended
super-wind driven by a highly-obscured star-burst, the wind then shocks
and ionizes the gas halo which surrounds the system.  This model can
explain several properties of these Ly$\alpha$ halos (although other
processes are also feasible, see Chapman et al.\ 2001 and Francis et
al.\ 2001) and hence deserves continued investigation, in particular
studying the kinematics of the extended Ly$\alpha$ emission around
SMM\,J17142+5016/\#18 using an integral-field spectrograph on an 8-m
class telescope to confirm the dynamical signatures of a super-wind.

Many of the characteristics of SMM\,J17142+5016/\#18 are shared with
other well-studied SCUBA galaxies (Ivison et al.\ 1998, 2000a): a
disturbed morphology, an SED dominated by dust-reprocessed radiation,
spectral signatures of a narrow-line AGN and extended Ly$\alpha$
emission. The disturbed appearance of \#18 sets it apart from the other
(typically-compact) Ly$\alpha$-selected galaxies and AGN around 53W002,
and suggests that, as in local ULIRGs, it is the tidal disturbance of
the gas reservoir within the galaxy due to a gravitational interactions
which is the root cause of the enhanced activity we see (Chapman et
al.\ 2002a).  Yamada et al.\ (2001) have argued that the
Ly$\alpha$-selected galaxies around 53W002 have characteristics similar
to those seen in the wider Lyman-break population, suggesting that
perhaps on-going major mergers are the events which set SCUBA galaxies
apart from the broader population of star-forming systems at high
redshifts. However, the presence of multiple components on 5--20\,kpc
scales in $\sim 30$\% of the Lyman-break population (Lowenthal et al.\
1997), along with the lack of significant submm emission from the
majority of Lyman-break galaxies (Chapman et al.\ 2000), indicate that
the picture is not that simple.

Although the exact trigger of the ultraluminous activity may be
ambiguous, it is also true that the power source for this behavior is
similarly difficult to identify.  The presence of an AGN in
SMM\,J17142+5016/\#18 obviously provides the opportunity for some
fraction of its far-infrared luminosity to come from AGN-heated dust
but, as with $z\sim 0$ ULIRGS, it is difficult to disentangle the
contributions from star formation and AGN-heating to the overall SED.
The existence of a large mass of dust, $10^8$\,M$_\odot$, suggests that
massive star formation has occurred in this galaxy, while the lack of
X-ray detections for the general SCUBA population (Almaini et al.\
2002), along with the constraints on AGN luminosities from a few
well-studied SCUBA galaxies (e.g.\ Frayer et al.\ 1998; Bautz et al.\
2000) both hint that only rarely does the AGN completely dominate the
bolometric output from typical, luminous submm-selected
sources. Sensitive {\it Chandra} X-ray observations will help to
resolve this issue for SMM\,J17142+5016/\#18, while {\it SIRTF}
mid-infrared observations will be key to disentangling the
contributions from the AGN to the far-infrared emission from this
galaxy.  More generally, it is clear from the frequency of weak AGN in
SCUBA galaxies that some of the gas from the large disturbed reservoirs
in these systems does find its way onto the central black hole and
given the easy availability of this fuel, the relatively modest
luminosity AGNs which result may indicate that the black holes are of
comparatively low mass (or suffer complex dust obscuration).

While the submm luminosity of SMM\,J17142+5016/\#18 sets it apart from
the typical Lyman break galaxy (Chapman et al.\ 2000), its relative
brightness in the optical/near-infrared wavebands also distinguishes it
from the other three submm sources in this field. All of these lack
obvious counterparts to $K_s\sim 20$ and are undetectable in our radio
map, although they should be detected close to the 3-$\sigma$ limit of
our radio map if they follow the mean Carilli \& Yun (2000) relation
and lie at $z\sim 2.4$.  Therefore, if these SCUBA galaxies inhabit the
same structure as SMM\,J17142+5016/\#18 they must have lower dust
temperatures (this will decrease their radio/far-infrared flux ratios,
making them fainter in the radio) and are also likely to be more
obscured.  We speculate that these characteristics could arise if these
submm sources lack the AGN which is seen in
SMM\,J17142+5016/\#18. Alternatively, several (or all) of these sources
could be unrelated to the structure around 53W002 and may lie at higher
redshifts. Unfortunately the limits placed on the sensitivity of the
radio map of this region by the presence of 53W002 means that improved
radio positions for these other submm sources are unlikely to become
available in the near future.

~From the current census of star formation activity in the structure
around 53W002 it appears that SMM\,J17142+5016/\#18 may be the
dominant contributor (at least compared to the Ly$\alpha$-selected
population).  We can thus estimate a crude lower limit to the star
formation density in this region, taking the rate calculated above and
assuming SMM\,J17142+5016/\#18 is the only ultraluminous galaxy in the
entire structure, which conservatively contains a volume of $4\times 4
\times 4$\,Mpc$^3$ (K99), giving a star formation density of $\gs
60$\,M$_\odot$\,yr$^{-1}$\,Mpc$^{-3}$.  Even allowing for a
significant contribution from the AGN to the bolometric luminosity,
this is still one to two orders of magnitude higher than the mean star
formation density at this epoch (Smail et al.\ 2002).  This is a
higher contrast than the 2--$3\times$ over-density of Ly$\alpha$
emitters in this region (several of which are also AGN, K99) and
suggests that vigorous, obscured star formation may be enhanced in the
protocluster environment.  

Finally we note that the confirmation of a luminous SCUBA galaxy
residing in a protocluster environment strengthens their associated
with the formation phase of the massive ellipticals, which dominate
rich clusters at the present-day.  The identification of this SCUBA
source adds an ultraluminous infrared galaxy to the diverse zoo of
galaxies inhabiting this structure at $z=2.4$, which now includes an
ultraluminous narrow-line AGN, a powerful radio galaxy, two broad-line
AGN and a large population of Ly$\alpha$-emitting galaxies.  If these
various types of activity are related through a single evolutionary
cycle, then environments such as that studied here may provide the
best insights into the details of their relationships.

\section{Conclusions}

We have identified the first, spectroscopically-confirmed SCUBA
companion to a high-redshift radio galaxy.  This is
SMM\,J17142+5016/\#18, which lies 330\,kpc ($42''$) and $\ls
300$\,km\,s$^{-1}$ offset in velocity from the steep-spectrum radio
galaxy 53W002 at $z=2.39$.  The restframe optical morphology of this
galaxy is strongly reminiscent of other well-studied SCUBA galaxies,
showing merger/tidal features (Ivison et al.\ 1998, 2000a).  We present
an OHS near-infrared spectrum of this galaxy, which probes the
restframe optical spectral properties, as well as a number of
long-wavelength constraints on the form of its SED. These show that
SMM\,J17142+5016/\#18 is a radio-quiet, narrow-line AGN with a
bolometric luminosity of $(8\pm 2)\times 10^{12}$\,L$_\odot$ and an
inferred star formation rate of $\sim 10^3$\,M$_\odot$\,yr$^{-1}$.  The
galaxy is also surrounded by an extended emission-line halo (seen in
Ly$\alpha$ and H$\alpha$).  These halos are being uncovered in an
increasing number of dusty, star-forming galaxies at high redshifts and
may signify the presence of starburst-driven super-winds in these
systems (Chapman et al.\ 2001, 2002b; Taniguchi \& Shioya 2000).  These
intense winds and the strong radiation fields produced by this activity
will have profound consequences for the evolution of any other galaxies
in close proximity to the SCUBA galaxies.

\acknowledgments

We thank Harald Kuntschner and Graham Smith for undertaking the INGRID
observations of this field, Fumihide Iwamuro and Toshinori Maihara for
obtaining the OHS spectra of SMM\,J17142+5016/\#18, Clare Jenner for
help with the SCUBA survey and Ian Waddington for the NICMOS imaging.
We are indebted to all staff members of the Subaru telescope, NAOJ, who
helped with the OHS observations.  We thank the referee for their quick
and concise report. We acknowledge useful conversations
with Andrew Blain, Scott Chapman, Len Cowie, Cedric Lacey
and Alice Shapley, as well as support from the Royal Society (IRS), the
Leverhulme Trust (DGG, IRS) and PPARC (JSD, JAS).

\end{document}